\documentclass[aps,prl,amsmath,amssymb,groupedaddress,superscriptaddress,floatfix,twocolumn,showkeys, amsart]{revtex4-2}

\usepackage[utf8]{inputenc}
\usepackage{xcolor}
\usepackage{relsize}

\usepackage{graphicx}
\usepackage{dcolumn}
\usepackage{mathtools}
\usepackage{makecell}

\newcommand{\sigsub}{\sigma_{\delta y}}
\newcommand{\sigsubduff}{\sigma_{\delta y}^{(D)}}
\newcommand{\sigyo}{\sigma_{y_0}}

\newcommand{\sigyosubduff}{\sigma_{\delta y_0}^{(D)}}

\newcommand{\phith}{\delta \phi_{\text{th}}}
\newcommand{\Ath}{\delta A_{\text{th}}}
\newcommand{\fth}{\delta f_{\text{th}}}
\newcommand{\fdrift}{\delta f_{\text{drift}}}
\newcommand{\fbeta}{\delta f_\beta}
\newcommand{\phibeta}{\delta \phi_\beta}
\newcommand{\Awhite}{\delta A_{\text{white}}}
\newcommand{\phiwhite}{\delta \phi_{\text{white}}}

\begin{document}
\title{High Stability Mechanical Frequency Sensing beyond the Linear Regime
}
 
\author{S. C. Brown}
\email{sofia.brown@colorado.edu}
\affiliation{JILA, National Institute of Standards and Technology and Department of Physics, University of Colorado, Boulder, CO 80309, USA}
\affiliation{Department of Physics, University of Colorado, Boulder, CO 80309, USA}
\author{R. Shaniv}
\altaffiliation{Current Address: Time and Frequency Division, National Institutes of Standards and Technology, Boulder, Colorado, 80305, USA}
\affiliation{JILA, National Institute of Standards and Technology and Department of Physics, University of Colorado, Boulder, CO 80309, USA}
\author{R. Zhang}
\affiliation{JILA, National Institute of Standards and Technology and Department of Physics, University of Colorado, Boulder, CO 80309, USA}
\affiliation{Department of Physics, University of Colorado, Boulder, CO 80309, USA}
\author{C. Reetz}
\affiliation{JILA, National Institute of Standards and Technology and Department of Physics, University of Colorado, Boulder, CO 80309, USA}
\author{C. A. Regal}
\affiliation{JILA, National Institute of Standards and Technology and Department of Physics, University of Colorado, Boulder, CO 80309, USA}
\affiliation{Department of Physics, University of Colorado, Boulder, CO 80309, USA}
\date{\today}

\begin{abstract}
Sensing a mechanical frequency shift is a powerful measurement tool. Therefore, understanding and mitigating frequency noise affecting mechanical resonators is imperative. The impact of noise on frequency sensing can be reduced with increased coherent amplitude of mechanical motion. However, large enough actuation places the resonator in the nonlinear (Duffing) regime, where conversion of amplitude noise into frequency noise can worsen sensor performance. Here, we present an experimentally straightforward method to evade this amplitude tradeoff in nano- or micro-mechanical sensors. Combining knowledge of the Duffing coefficients with readily available amplitude measurements, we avoid amplitude-to-frequency noise conversion. We use dual-mechanical-mode operation on a tensioned thin-film resonator to set a baseline thermomechanically-limited stability by eliminating correlated single-mode frequency drifts. Thus, we observe amplitude-to-frequency noise conversion at high drive and reduce it using our method. The resulting high-stability operation beyond the linear regime contrasts long-standing perspectives in the field. 
\end{abstract}

\maketitle

The resonance frequency of a mechanical mode can be used to probe the resonator's environment. For example, a mechanical resonance frequency may change due to a change in material stress, environmental temperature, electric and magnetic fields, or mass. Frequency detection therefore enables sensing technologies such as mass sensing \cite{ekinci2004ultrasensitive, yang2006zeptogram, hanay2012single}, spin sensing \cite{giessibl2003advances, poggio2010force, rugar2004single}, and thermal bolometric sensing \cite{li2023terahertz, blaikie2019fast, vicarelli2022micromechanical, piller2022thermal, zhang2024high, zhang2025enhanced, luhmann2023nanoelectromechanical, martini2025uncooled}. For many applications, nano- and micromechanical resonators, with dimensions on the order of hundreds of nanometers to hundreds of microns, are particularly suitable, due to their high frequency and small footprint and the ability to fabricate them reliably and detect their motion precisely. Such devices are important for nanoscale sensing, where a resonance frequency shift arises from, e.g., the presence \cite{ekinci2004ultrasensitive, yang2006zeptogram} or thermal desorption \cite{piller2022thermal} of a nanoparticle or complex molecules on the resonator.

Typically, the mechanical resonator is designed and functionalized to enhance sensitivity to one parameter while minimizing sensitivity to others, which are considered noise. To improve sensor resolution and bandwidth, continued effort has been put into understanding and mitigating frequency and phase noise \cite{sansa2016frequency, gavartin2013stabilization, kharbanda2025chip, fong2014phase, maillet2016classical, maillet2018measuring, manzaneque2023resolution}.

The fundamental limit to mechanical frequency detection is thermomechanical noise, resulting from Brownian motion. This motion imprints random noise on the amplitude and phase quadratures of the resonator response, and we hereafter refer to these noises as $\Ath$ and $\phith$, respectively. The temporal derivative of  $\phith$ causes a detected angular frequency noise, $2\pi\fth$.

The thermomechanical limit can be reduced through mechanical driving to high coherent amplitude. However, the trend of higher amplitude leading to better frequency stability often relies on the resonator being in its linear regime \cite{manzaneque2023resolution, schmid2016fundamentals, demir2019fundamental, roy2018improving}. 
At high amplitudes, resonator dynamics are influenced by Duffing nonlinearity \cite{schmid2016fundamentals}. In this regime, nonlinear mixing between amplitude and frequency causes any noise on the amplitude, $\delta A$, e.g.~thermomechanical amplitude noise \cite{maillet2017nonlinear} or noise on the external drive, to be transduced to frequency noise, $\delta f$. 

To optimize the tradeoff between high amplitude operation and $\delta A$-$\delta f$ conversion, conventional guidance is often to operate at the critical amplitude, where nonlinearity sets in \cite{kanellopulos2025comparative, olcum2014weighing, molina2021high, schmid2016fundamentals, sansa2016frequency, demir2019fundamental, roy2018improving}. However, this approach prevents full use of a high mechanical drive and, in smaller devices, the nonlinearity presents at lower amplitudes \cite{barnard2019real, barnard2012fluctuation, eichler2013symmetry, maillet2017nonlinear}, limiting obtainable phase resolution. Some suggest that choosing between optimal performance in either the slow or fast sensing regime should determine whether to drive beyond linearity  \cite{manzaneque2023resolution}. Others have developed ways to overcome $\delta A$-$\delta f$ conversion at high amplitude \cite{zou2017non, kenig2012optimal, villanueva2013surpassing, antonio2012frequency, zhang2024frequency}, but they are complicated experimentally and/or functional only in a limited parameter space. 

In this Letter, we present and demonstrate a robust approach to operate above the critical amplitude, contrary to the belief that such operation always degrades performance. Using the simple and widespread resonance frequency detection platform, the phase-locked loop (PLL) \cite{schmid2016fundamentals, demir2019fundamental, bevsic2023schemes, olcum2015high}, our method leverages readily available but typically unused time-dependent amplitude information to correct $\delta A$-$\delta f$ conversion. We thus improve the stability of our resonator by an order of magnitude when driven well beyond the critical amplitude. Furthermore, using a multi-mechanical-mode detection scheme on our 120 nm-thick patterned thin-film resonator \cite{norte2016mechanical}, we mitigate ambient temperature drifts. Thus, we demonstrate correctable $\delta A$-$\delta f$ effects and reach thermomechanically-limited noise levels below $5\times10^{-10}$ fractional frequency Allan deviation \cite{sadeghi2020frequency, kanellopulos2025comparative, zhang2025enhanced, sansa2016frequency}, independent of chosen operating parameters. 

We consider an oscillator where the leading nonlinearity is the Duffing nonlinearity, a geometric hardening effect in which the resonator stiffness changes with displacement, no longer in accordance with Hooke's Law. For a single damped and driven Duffing oscillator with resonant frequency $\omega_0$, quality factor $Q$, sinusoidal time-dependent driving force $F(t)$, and Duffing parameter $\beta$, the equation of motion for the oscillator position in time, $z(t)$, is given by

\begin{equation}
    \ddot z(t) + \frac{\omega_0(t)}{Q}\dot z(t) + \omega_{0}^2(t) \left(1 +  \frac{\beta}{\omega_{0}(t)^2} z^2(t) \right) z(t) = \frac{F(t)}{m}.
    \label{Eq: Duffing EOM}
\end{equation}
From Eq.~\ref{Eq: Duffing EOM}, it is apparent that oscillator's frequency depends on $z^2$. 

The oscillatory motion of the resonator can be described in terms of a time-dependent amplitude, $A(t)$, and phase, $\phi(t)$, i.e. $z(t) = A(t) \sin (\omega_0(t)t + \phi(t))$. To measure a resonator’s frequency, $\omega_0(t)$, an external sinusoidal driving force is applied, and the oscillator response is recorded. The oscillator's amplitude and phase are then obtained in the force's rotating frame via demodulation. The phase is fed to a PLL, locking the driving frequency to the mechanical resonance and making it the detected quantity. 

\begin{figure}[t]
    \centering
    \includegraphics{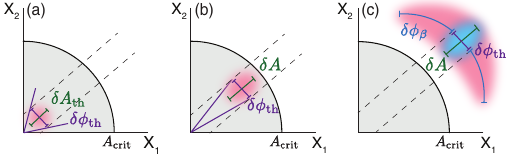}
    \caption{
    Phase space plots illustrating the Duffing problem and our solution. (a) At a low drive, the resonator amplitude (green lines) and phase inference (purple lines) are dominated by thermomechanical noise. (b) Increasing the amplitude, while remaining below $A_{\text{crit}}$ decreases the relative contribution of $\phith$ to the resonator phase inference (angle created by purple lines is smaller in (b) than in (a)). Other noises in addition to thermomechanical noise can be present in the amplitude, leading to a total amplitude noise, $\delta A$. (c) Driving beyond $A_{\text{crit}}$ leads to $\delta A$-$\delta f$ conversion (red) and thus a detected phase noise due to Duffing effects, $\phibeta$. Our method removes $\fbeta$, resulting in the blue ellipse.} 
    \label{Fig: Duffing subtraction concept}
\end{figure}

Because this detection method treats amplitude and phase independently, it allows arbitrary choice of drive amplitude. As shown in the phase space of Fig.~\ref{Fig: Duffing subtraction concept}(a), (b), when limited by thermomechanical noise and operating in the linear regime, a larger coherent amplitude resolves the changing phase (frequency) signal above the stochastic thermal phase noise \cite{cleland2002noise}. Here, X$_1$ and X$_2$ are the demodulated quadratures of the resonator motion. Thus, in this case, driving with the highest amplitude possible is beneficial.

Eq.~\ref{Eq: Duffing EOM} suggests, however, that beyond a certain critical amplitude, $A_\text{crit}$ (black arcs in Fig.~\ref{Fig: Duffing subtraction concept}), the effect of the nonlinearity on the resonator frequency becomes significant compared to the linewidth of the mechanical mode. For a single Duffing oscillator, this amplitude is given by \cite{schmid2016fundamentals}

\begin{equation}
    A_{\text{crit}} = \sqrt{\frac{8}{3\sqrt{3}}}\frac{1}{\sqrt{Q}}\sqrt{\frac{\omega_0^2}{\beta}}.
    \label{Eq: critical amp}
\end{equation}
Beyond $A_{\text{crit}}$, the oscillator amplitude and frequency are coupled such that any amplitude noise, $\delta A = A-\langle A \rangle$, manifests as frequency noise, $2 \pi\fbeta \approx \beta \times (\delta A(t))^2$. This frequency noise results in a corresponding phase noise, $\phibeta(t)$, detected by the PLL, where $\phibeta(t) \sim \int \beta \times (\delta A(t))^2dt$ (red noise profile in Fig.~\ref{Fig: Duffing subtraction concept}(c)). Typically $\delta A > \Ath$ when the resonator is driven and may change with drive amplitude. In this work, we develop a method to remove $\fbeta$ (blue ellipse in Fig.~\ref{Fig: Duffing subtraction concept}(c)) regardless of the character of $\delta A$.

To characterize the frequency noise present in a system, the standard metric is the fractional frequency Allan deviation, $\sigma_y(\tau)$, which quantifies frequency stability at a given averaging time $\tau$ \cite{schmid2016fundamentals, demir2019fundamental, Allan1986}. Here, $y =\frac{\delta f}{\langle f \rangle} = \frac{f- \langle f\rangle}{\langle f \rangle}$. $f$ is the resonator frequency as detected by the PLL, and $\langle f\rangle$ is the average frequency of the data trace. 

The aforementioned thermomechanical frequency noise arises from a white thermal force that leads to resonator phase fluctuations. By analyzing the power spectral density (PSD) associated with these fluctuations, a fractional frequency Allan deviation, $\sigyo$, which characterizes the asymptotic frequency stability of a system limited by $\fth$, can be obtained \cite{manzaneque2023resolution, demir2019fundamental, bevsic2023schemes}: 
\begin{equation}
   \sigyo = \sqrt{\frac{k_B T}{m\omega_0^3Q}} \times \frac{1}{A_0\sqrt{\tau}},
   \label{Eq: Thermomech limit}
\end{equation}
where $m$ is the mode's effective mass \cite{schmid2016fundamentals}, $k_B$ is the Boltzmann constant, and $A_0$ is the average amplitude to which the oscillator is driven, i.e. $A_0 = \langle A(t) \rangle$. Notably, $\sigyo$ depends inversely on $A_0$. It is also inversely proportional to $\sqrt{\tau}$, as expected for a white frequency noise source.

Historically, the thermomechanical frequency limit has been difficult to reach in nano- and micro-mechanical resonators even in the linear regime due to many noise sources including imprecision noise, bulk and surface effects, molecular adsorption and desorption, dielectric and charge fluctuations, and two-level-system (TLS) noise \cite{sansa2016frequency, fong2014phase}. In particular, reaching this limit often requires at minimum overcoming long-term technical frequency drifts, $\fdrift$, on the scale of seconds to thousands of seconds arising from, e.g., changes in the overall resonator temperature \cite{gavartin2013stabilization, sadeghi2020frequency, kharbanda2025chip}. Drift frequency noise typically manifests as a worsening of the Allan deviation at long times with a power of $\tau$ greater than zero.

\begin{figure*}[t!]
    \centering
    \includegraphics{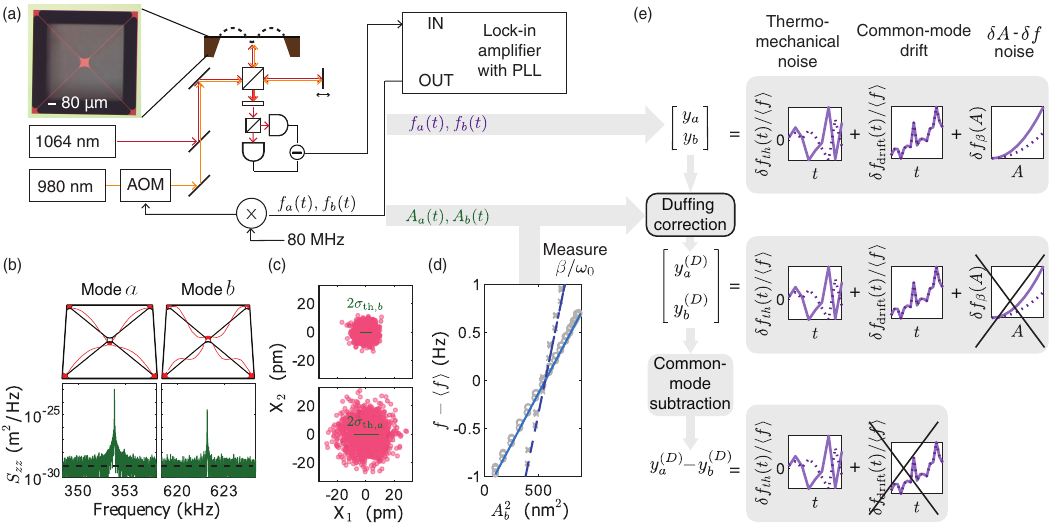}
    \caption{Experimental setup and methods. (a) Optical drive and interferometric detection of two modes of a tensioned Si$_3$N$_4$ trampoline (microscope image). Crucially, we record both modes' frequencies \textit{and} amplitudes, in time. A high-pass optical filter prevents 980 nm light from reaching the detector. (b) Undriven amplitude noise PSD of modes $a$ and $b$ with respective mode shapes above (flat resonator in black, and out-of-plane displacement of each mode in red). Black dashed lines show the imprecision noise floor. (c) Undriven phase space plots for each mode (mode $a$ below, mode $b$ above). The distributions agree with thermal expectations (green lines). (d) Example of two of the four Duffing-coefficient measurements. Detunings of each mode vs.~$A_b^2$ as the driving force is turned on, with arbitrary y-intercepts. Lines (solid light blue for mode $a$ frequency, dashed dark blue for mode $b$ frequency) represent the average of many linear fits of data from multiple cycles of turning on the driving force and reaching steady-state oscillation (see S.2.2) (circles for mode $a$ and x for mode $b$). e) Conceptual representation of the dual-mode enabled Duffing correction method. Noise contributions of mode $a$ (solid light purple) and of mode $b$ (dotted dark purple) to the relevant frequency signals are shown symbolically as plots versus either time or amplitude as our method is employed. 
    }
    \label{Fig: Setup}
\end{figure*}

Fig.~\ref{Fig: Setup} explains how our method both achieves high stability in the nonlinear regime over a large bandwidth, unencumbered by $\delta A$-$\delta f$ conversion, and reaches this stability despite drifts. Our method employs two mechanical frequencies, $f_a$ and $f_b$, of modes existing on a suspended, patterned, 2D membrane. This platform is well-suited to this multi-mechanical-mode operation as it allows designing an array of modes with varying properties. For the Duffing correction, we combine the modal amplitude, with its associated noise, $\delta A$, and the resonator Duffing coefficients to negate the $\fbeta$ of each mode (Fig.~\ref{Fig: Setup}(e)). The amplitude is directly measured in the PLL frequency tracking scheme as a readily available but typically unused information in similar experiments. Since $\beta$ is a material and geometric property of the resonator, it is easily calibrated beforehand (Fig.~\ref{Fig: Setup}(d)) \cite{catalini2020soft, catalini2021modeling}.

With $\fbeta$ reduced, the frequency of interest is the common-mode rejection between the two modes. This subtraction suppresses drifts fractionally common to both modes \cite{gavartin2013stabilization, kharbanda2025chip,wang2000temperature} and establishes a baseline stability limited by $\fth$, without temperature control or enclosure of the device (Fig.~\ref{Fig: Setup}(e)). We emphasize that the Duffing correction is applicable independently of the common-mode rejection. 

For dual-mode operation, we must consider not only the effect of a given mode's nonlinearity on its own frequency (self-Duffing), but also its effect on the other mode's frequency (cross-Duffing) \cite{westra2010nonlinear}. This coupling between mode $i$ and $j$ produces the following Duffing equation of motion for mode $i$, identical to Eq.~\ref{Eq: Duffing EOM} but with a cross term:
\begin{equation}
\begin{split}
&\frac{F(t)}{m_i}=\\
&\ddot z_i(t) +\frac{\omega_{0,i}}{Q_i}\dot z_i(t) + \omega_{0,i}^2 \left (1 + \frac{\beta_{ii}^{\text{SD}}}{\omega_{0,i}^2} z_i^2(t) + \frac{\beta_{ij}^{\text{XD}}}{\omega_{0,i}^2}z_j^2(t)\right) z_i(t), 
    \label{Eq: EOM}
    \end{split}
\end{equation}
where, following notation in \cite{catalini2020soft}, $\beta^{\text{SD}}_{ii}$ is now the self-Duffing parameter, and $\beta^{\text{XD}}_{ij}$ is the cross-Duffing parameter.

By solving Eq.~\ref{Eq: EOM} \cite{catalini2020soft}, we find that the instantaneous frequency shift due to the Duffing nonlinearity in dual-mode operation is

\begin{equation}
    f_{0,i}' - f_{0,i} = \gamma_{ii}^{\text{SD}} A_i^2 + \gamma_{ij}^{\text{XD}} A_j^2, 
    \label{Eq: Duffing Freq Shift}
\end{equation}
where $f_{0,i}(t) = \frac{\omega_{0,i}(t)}{2 \pi}$ and each $\gamma$ is proportional to its respective $\beta$ divided by $\omega_{0, i}$.

The expression for the critical amplitude is more complicated when considering two modes, as it must depend not only on $\beta_{ii}^{\text{SD}}$ but also on $\beta_{ij}^{\text{XD}}$. In this work, we refer only to the self-Duffing critical amplitude \cite{schmid2016fundamentals}, $A_{\text{crit}}^{\text{SD}}$, defined in Eq.~\ref{Eq: critical amp} (with $\beta = \beta_{ii}^{\text{SD}}$) when we require a scale for the onset of nonlinearity.

Our device is a 1.5 mm tensioned trampoline resonator \cite{norte2016mechanical} made with 120 nm thick Si$_3$N$_4$ (microscope image in Fig.~\ref{Fig: Setup}(a), fabrication details in S.1). The resonator resides in a vacuum chamber at $4.5\times10^{-8}$ Torr to avoid gas damping. The resonator displacement carries all the information needed in the experiment and is read out with a 1064 nm YAG laser beam via a calibrated Michelson interferometer (Fig.~\ref{Fig: Setup}(a)). The interferometer local oscillator arm length is locked through the feedback-controlled pizeo mirror position. The imprecision noise floor of our detection laser system is a few fm/$\sqrt{\text{Hz}}$ (dashed lines in Fig.~\ref{Fig: Setup}(b)), which allows our modes' thermomechanical motion to be resolved more than five orders of magnitude above this noise (S.3.1).

For the dual-mode operation, we choose two higher-order, symmetric trampoline modes (S.3.2). Mode $a$ has $f_{0, a}=352$ kHz and mode $b$ has $f_{0, b} = 622$ kHz (Fig.~\ref{Fig: Setup}(b)). To verify that our undriven motion is thermal, we measure the two quadratures of the undriven motion of each mode and compare to theory (Fig.~\ref{Fig: Setup}(c)). As expected, the fluctuations are Gaussian in both quadratures with a standard deviation, $\sigma_{\text{th}}$, matching that predicted by the equipartition theorem, $\sigma_{\text{th}} = \sqrt{\langle z_{\text{th}}^2 \rangle} = \sqrt{\frac{k_B T}{m \omega_o^2}}$, where $T =$ 295 K and $m$ comes from COMSOL finite element analysis simulation, assuming displacement detection at the trampoline center. 

Our two modes are driven optically via radiation pressure using an amplitude-modulated 980 nm diode laser (Fig.~\ref{Fig: Setup}(a), S.4). We use tens of $\mu$W of modulated driving power to achieve amplitudes of tens of nm. A double-demodulator lock-in amplifier (Zurich Instruments, MFLI) with two PLLs simultaneously tracks the frequencies of both modes, in time, by comparing the phase of the driving beam modulation, output by the lock-in, to that of the interferometric response. Crucially, the amplitudes of each mode, $A_a(t)$ and $A_b(t)$, are also recorded by the demodulators. 

We use a 500 Hz demodulator filter bandwidth for both modes. PI parameters are chosen to yield a frequency-response bandwidth of $\sim 50$ Hz for both modes and therefore an observed Allan deviation independent of these parameters for $\tau > 0.2$ s \cite{olcum2015high, demir2021understanding, bevsic2023schemes}. To demonstrate the simplicity and robustness of our method, we maintain constant PLL lock parameters and the setpoint fixed at the $\pi/2$ phase locking point throughout all closed-loop measurements, regardless of drive amplitude. Even without PLL phase optimization \cite{cuairan2022precision, kenig2012optimal, villanueva2013surpassing}, we achieve high frequency stability implying that PLL bistability is not a significant effect.

\begin{figure}[t!]
    \centering
    \includegraphics{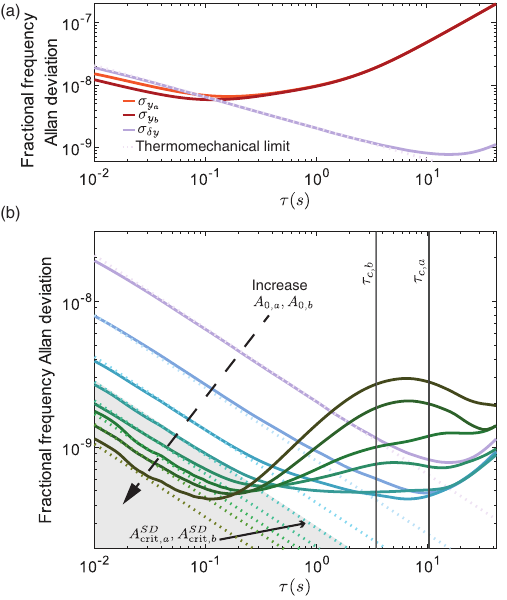}
    \caption{Common-mode subtraction data and theory for different drive amplitudes. Solid lines are measured data and dotted lines are the room-temperature thermomechanical limit (Eq.~\ref{Eq: Thermomech limit}) of both modes added in quadrature for each drive. (a) $\sigma_{y_a}$ and $\sigma_{y_b}$  (orange and red) and corresponding $\sigsub$ (light purple) for $A_{0,a}=$ 1.4 nm and $A_{0,b} =$ 0.71 nm. Purple lines are identical to lowest amplitude solid and dotted line in (b) Measured $\sigsub$ for increasing drive amplitude, where $\frac{A_a}{A_{\text{crit},a}^{\text{SD}}} = \frac{A_b}{A_{\text{crit},b}^{\text{SD}}} =  $ 0.14, 0.35, 0.71, 1.06, 1.41, 1.77, 2.12, 2.83. Here, $A_{\text{crit},a}^{SD} \approx 10$ nm, $A_{\text{crit},b}^{SD} \approx 6$ nm.
    }
    \label{Fig: Subtracted ADevs with Theory}
\end{figure}

\begin{figure*}[t!]
    \centering
    \includegraphics[scale = 1.1]{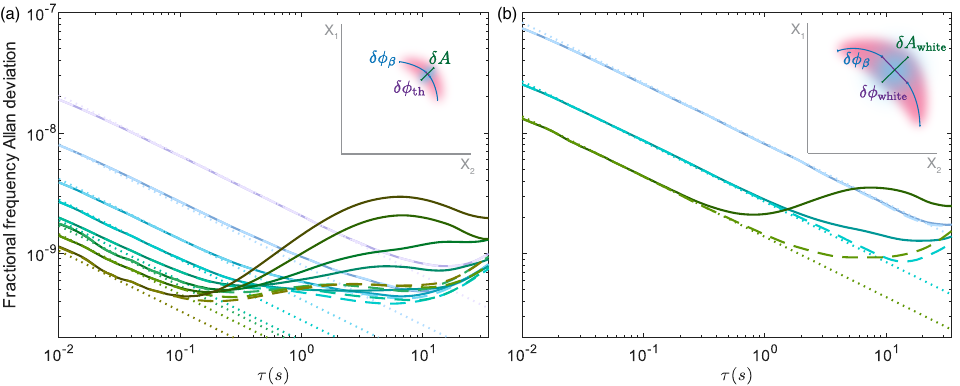}
    \caption{Duffing correction results. (a) $\sigsub$ (solid lines) and $\sigsubduff$ (dashed lines) for the same drive amplitudes as in Fig.~\ref{Fig: Subtracted ADevs with Theory}(b) and corresponding thermomechanical limits (dotted lines). (b) Results including added white noise that elevates the effective temperature. Shown are three example traces of $\sigsub$ (solid lines) and $\sigsubduff$ (dashed lines) with colors as described in Fig.~\ref{Fig: Subtracted ADevs with Theory}(b). Here the elevated effective noise values (dotted lines) are obtained using a free parameter fit. (Insets) Illustrative phase space diagrams without and with added noise (see Fig.~\ref{Fig: Duffing subtraction concept}) for a hypothetical actuation above the critical amplitude for before (red noise profile) and after (blue noise profile) Duffing correction.}
    \label{Fig: Duffing Correction}
\end{figure*}

To characterize our Duffing cancellation, we define $y_i = \frac{\delta f_i}{\langle f_i \rangle}$ and the Allan deviation of this quantity as the single-mode Allan deviation, $\sigma_{y_i}$. For the common-mode rejection, we define $\sigsub$ to be the Allan deviation of the quantity $y_a -y_b$. To then correct $\sigsub$ for Duffing $\delta A$-$\delta f$ conversion as per Eq.~\ref{Eq: Duffing Freq Shift}, we define

\begin{equation}
y_i^{(D)} = \frac{f_i - \gamma^{\text{SD}}_{ii}A_i^2 - \gamma^{\text{XD}}_{ij}A_j^2}{\langle f_i - \gamma^{\text{SD}}_{ii}
A_i^2 - \gamma^{\text{XD}}_{ij}A_j^2\rangle},
\end{equation}
such that we can define $\sigsubduff$, the Duffing correction, as the Allan deviation of 

\begin{equation}
y_a^{(D)}-y_b^{(D)}.
\label{Duffing Correction}
\end{equation}
Here, we follow Refs. \cite{catalini2020soft} and \cite{catalini2021modeling} to experimentally determine $\gamma_{ii}^{\text{SD}}$ and $\gamma_{ij}^{\text{XD}}$ (Fig.~\ref{Fig: Setup}(d) and Table S.1). We turn on the drive of mode $i$ and fit $f_i$ and $f_j$ vs.~$A_i^2$ to lines according to Eq.~\ref{Eq: Duffing Freq Shift} as the amplitude of mode $i$ increases (see S.2). In this work, $y_i^{(D)}$ is computed in post-processing (Fig.~\ref{Fig: Setup}(e)), but this correction may be extended to real time.

Fig.~\ref{Fig: Subtracted ADevs with Theory}(a) shows the single-mode Allan deviations for modes $a$ and $b$, $\sigma_{y_a}$ and $\sigma_{y_b}$, and the corresponding $\sigsub$, for a low drive of $A_{0,a} = 1.4$ nm, $A_{0,b} = 0.71$ nm. Clearly, $\sigma_{y_a}$ and $\sigma_{y_b}$ are dominated by $\fdrift$ at long times. However, the drifts are fractionally correlated and are well subtracted (purple solid line) (S.3). The dotted line shows the thermomechanical limit, calculated from Eq.~\ref{Eq: Thermomech limit}, of the two modes added in quadrature, assuming room temperature. When operating firmly in the regime, $\sigsub$ reaches this limit up to $\tau \sim 10$ s.

Fig.~\ref{Fig: Subtracted ADevs with Theory}(b) shows $\sigsub$ traces for increasing drive amplitudes of both modes, where for each trace, $A_{0, a}/A_{\text{crit}, a}^{\text{SD}} = A_{0, b}/A_{\text{crit}, b}^{\text{SD}}$. For all amplitudes at $\tau < 0.1$ s, $\sigsub$ reaches the expected thermomechanical limit without a free-parameter fit. For low drive amplitudes, $\sigsub$ continues to reach this limit at longer times, but beyond $A/A_{\text{crit}}^{\text{SD}} \approx 1$ for both modes, the expected tradeoff between improved sensitivity and $\delta A$-$\delta f$ conversion is apparent in the local maxima occurring around $\tau_{c,a}$ and $\tau_{c,b}$, the characteristic timescales of the Duffing noise, defined as $\tau_c = \frac{2\sqrt{3}Q}{\omega_0}$ \cite{manzaneque2023resolution}. The minimum Allan deviation of the highest amplitude trace (darkest green) matches the value expected of a Duffing oscillator, $\sigma_{y,min}(\tau << \tau_c) = \frac{\sqrt[4]{3}}{2}\sqrt{\frac{k_BT\gamma^{SD}}{Qm\omega_0^2}}$ \cite{manzaneque2023resolution}, when $\sigma_{y,min, a}$ and $\sigma_{y,min, b}$ are added in quadrature.

We now show that this $\delta A$-$\delta f$ conversion can be significantly removed. Fig.~\ref{Fig: Duffing Correction}(a) shows $\sigsub$ (solid lines) and $\sigsubduff$ (dashed lines) for the same drive amplitudes as in Fig.~\ref{Fig: Subtracted ADevs with Theory}. $\sigsubduff$ traces at high drive amplitude show higher stability over a larger bandwidth than $\sigsub$ traces at linear drive amplitudes. In Fig.~\ref{Fig: Duffing Correction}(a), we also show a phase space noise profile for an example actuation level in the nonlinear regime without (red) and with (blue) the Duffing correction. We find that the amplitude noise $\delta A$ contributing to $\fbeta$ is highly drive-dependent and produces $\delta A$-$\delta f$ conversion an order magnitude larger than that expected from the $\Ath$ of both modes alone (S.5). 

The Duffing correction does not reach the thermomechanical frequency limit (dotted lines) at high drive for $\tau > 1$ s. Indeed, in this regime of high stability, the Duffing correction allows us to observe residual limiting frequency noise, unattributable to Duffing effects. The flat behavior of the Duffing correction at long times suggests limiting $1/f$ noise, a noise spectrum common to similar experiments and minimally understood. Cited sources of this noise include defect motion with a broad distribution of relaxation times and related (TLS-) like models and damping fluctuations \cite{sansa2016frequency, fong2014phase, maillet2018measuring}.

We also analyze the efficacy of the Duffing correction in a regime where the amplitude and phase noise are dominated by white noise, $\Awhite$ and $\phiwhite$ (Fig.~\ref{Fig: Duffing Correction}(b)), rather than the residual noises just described. We add white force noise to our radiation pressure drive, effectively simulating elevated thermomechanical noise \cite{manzaneque2023resolution} at 27,000 K (see S.4.1 for calibration). As shown in the phase space diagram of Fig.\ref{Fig: Duffing Correction}(b), the amplitude noise is now dominated by this added $\Awhite$, although some extraneous drive-dependent amplitude noise remains at high drive (S.5). Additionally, the magnitude of phase noise resulting from white force noise is increased by a factor of 10 such that $\phiwhite$ now dominates over the residual $1/f$ noise. In this regime, $\sigsubduff$ continues to improve with averaging time as $1/\sqrt{\tau}$, leading to two conclusions: first, that in the ideal case where residual noises are absent, averaging longer under our method leads to strictly improved sensor performance and second, that our method corrects for $\delta A$-$\delta f$ conversion arising even from noise with the character of fundamental thermomechanical amplitude noise.

In conclusion, we have demonstrated mechanical frequency stability that remains high over long averaging times even in the Duffing regime. Our method of common-mode subtraction and Duffing correction is robust, uses information accessible in a closed-loop frequency-tracking setup, and can be extended to real time processing, if desired. Our method enables broader investigation of noises in micromechanical resonators that appear at instability less than $10^{-9}$ Allan deviation. In the future, the $1/f$ behavior limiting the Duffing correction may be studied. We demonstrate that our method applies regardless of the source of $\delta A$–$\delta f$ conversion, indicating that even in ultrahigh-stability devices, conversion from fundamental thermomechanical noise is removed. 

Our platform of a tensioned membrane facilitates multi-mode operation, given the ease of multi-mode design and manipulation in this space \cite{tsaturyan2017ultracoherent, reetz2019analysis, norte2016mechanical, shin2022spiderweb, fedorov2020fractal}. It opens the design space for a high-performance micromechanical bolometer \cite{li2023terahertz, blaikie2019fast, vicarelli2022micromechanical, piller2022thermal, zhang2024high, zhang2025enhanced, luhmann2023nanoelectromechanical, martini2025uncooled} that uses our noise reduction method. For example, when a device is functionalized with a thermal absorber, two modes with a differential response to localized heat but a common-mode response to technical drifts could reduce sensor noise while maintaining a clear signal. Another salient topic is to evaluate our method on other resonator geometries and types. Lower mass resonators, where the nonlinear regime is more quickly accessed, are of particular interest.

Acknowledgments: The authors thank Peter Steenken, Silvan Schmid, Scott Papp, and Jack Harris for insightful discussions; Luca Talamo, Sarah Dickson, Michelle Chalupnik, Maxwell Urmey, and Kazemi Adachi for experimental assistance and careful reading of the manuscript; and David Carlson for assistance with the fabrication. This work was supported by funding from AFOSR under award number FA9550-24-1-0173, NSF Grants PHY-2317149 and NSF QLCI award OMA-2016244, the Brown Institue for Basic Sciences, and the Baur-SPIE Endowed Chair at JILA.  

\appendix
\setcounter{secnumdepth}{3}
\renewcommand{\thesection}{\Alph{section}}
\section{Device Fabrication}
\label{Section: Fab}

\indent The device used in this work is a tensioned Si$_3$N$_4$ 1.5 mm trampoline resonator, with an 80 $\mu$m wide pad and 5 $\mu$m wide tethers (specifications in Table \ref{Table: Device Params}). The Si$_3$N$_4$ was deposited on silicon using LPCVD for 53 minutes at 150 mTorr with a gas ratio of 75/25 sccm for NH$_3$/DCS. The deposition was done without annealing and resulted in an intrinsic stress of $\sim$1 GPa and a Si$_3$N$_4$ thickness of 120 nm. We performed photolithography to pattern the device. Exposed regions of Si$_3$N$_4$ were then etched away using reactive ion etching, and the underlying Si was removed through a KOH etch such that the remaining Si$_3$N$_4$ trampoline was left suspended over a window.

\begin{figure}[t!]
    \centering
    \includegraphics{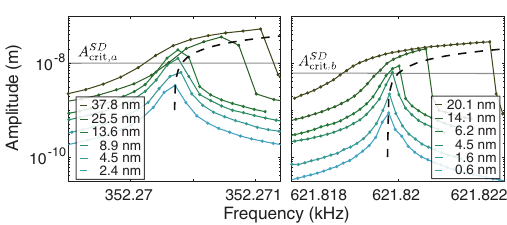}
    \caption{\small Open-loop amplitude responses of modes $a$ (left) and $b$ (right) as the drive frequency is swept across resonance from low to high frequency. Different colored traces show difference drive amplitudes as labeled in nm. The dotted black lines indicate the theoretical backbone curve of each mode.}
    \label{Fig: BackBoneCurves}
\end{figure}

\begin{table*}[t]
\centering
\fontsize{9pt}{10pt}\selectfont

\begin{tabular}{|c|c|c|c|c|c|c|c|}
\hline
&
\makecell{\textbf{Resonance}\\\textbf{frequency} \\$f_0$} &
\makecell{\textbf{Quality}\\\textbf{factor}  \\ $Q $} &
\makecell{\textbf{Effective}\\\textbf{mass} \\$m$} &
\makecell{\textbf{Critical}\\\textbf{amplitude} \\$A_{\text{crit}}^{\text{SD}}$} &
\makecell{\textbf{Self-Duffing}\\\textbf{coefficient} \\ $\gamma^{\text{SD}}$} &
\makecell{\textbf{Cross-Duffing}\\\textbf{coefficient} \\$\gamma^{\text{XD}}$} &
\makecell{\textbf{Duffing}\\\textbf{time constant} \\$\tau_c$} \\
\hline
Units& kHz & 10$^6$ & ng & nm & Hz/nm$^2$ & Hz/nm$^2$ & s\\ \hline
Mode $a$ & 352.2 & 7 & 10.7 & 10 & $5.93 \times 10^{-4}$ & $2.09 \times 10^{-3}$ & 10.4  \\
Mode $b$ & 621.7 & 4 & 17.7 & 6 & $5.44 \times 10^{-3}$ & $7.06 \times 10^{-4}$  & 3.5 \\
\hline
\end{tabular}
\caption{Trampoline resonator parameters. $Q$ is obtained from ring-down measurements. The critical amplitude listed considers only self-Duffing effects and provides a scale of the onset of nonlinearity.}
\label{Table: Device Params}
\end{table*}

\section{Duffing Behavior of Modes a and b}
\subsection{Open-Loop Amplitude-Frequency Responses}

Fig.~\ref{Fig: BackBoneCurves} shows the open-loop response of modes $a$ and $b$ as the drive frequency is swept across resonance from low to high frequency. Increasing drive strengths are shown as different colored traces. Due to the high Q of the two modes, drifts in frequency greater than the modal linewidths were observed between consecutive sweeps.  Because these drifts are largely correlated (see \ref{Sec: Correlated Drifts}), we track the frequency of mode $j$ in the linear regime while sweeping mode $i$ and measuring mode $i$'s response. We then use the measured fractional drifts of mode $j$ to correct for drifts of mode $i$ during sweeps.

The expected transition between a linear Lorentzian and nonlinear response is clearly visible as the drive amplitude is increased. This behavior is consistent with the expected backbone curve (dotted lines) describing the resonance frequency shift due to Duffing effects given by \cite{catalini2020soft}
\begin{equation}
    \omega_{max}  = \omega_0 +\gamma^{SD}A_{max}^2,
\end{equation}
where $\omega_{max}$ is the resonance frequency at the maximum amplitude $A_{max}$, and $\gamma^{SD}$ is the measured self-Duffing coefficient (see \ref{DuffingCalib}). The transition from a linear to a Duffing response also occurs around the expected $A^{SD}_{crit,a} \approx$ 10 nm and $A^{SD}_{crit,b} \approx$ 6 nm (horizontal gray lines). Deviations between the data and backbone curve may be partially attributed to the small linewidth and the drift correction, and we note that the Duffing measurement described in \ref{DuffingCalib} provides the accurate measurement of our Duffing coefficient used in our technique.

\subsection{Duffing Coefficient Calibration}
\label{DuffingCalib}

\begin{figure}[t!]
    \centering
    \includegraphics{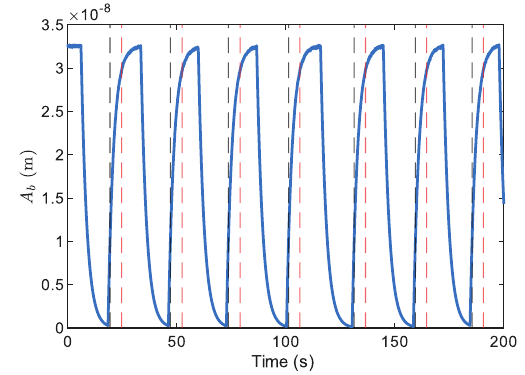}
    \caption{\small Example data used for Duffing coefficient calibration as the optical drive is turned on and off. The amplitude of mode $b$ in time displays multiple ring ups (excitations) and ring downs (decays). Black (red) dotted lines indicate the beginning (end) of data in a given ring up cycle used to extract the Duffing coefficient.}
    \label{Fig: DuffingInTime}
\end{figure}

We now explain in detail how the four Duffing coefficients, $\gamma$, are measured. With both modes' PLLs engaged, we turn on and off the optical drive of mode $i$ approximately 20 times, with a period of the mode's mechanical time constant. Thus, we achieve multiple ring ups (excitations) and ring downs (decays) of mode $i$'s amplitude (Fig.~\ref{Fig: DuffingInTime}). During these excitation cycles, we separately record the frequencies of modes $i$ and $j$ in time. We then select the frequency data corresponding to ring ups, i.e.~data corresponding to mode $i$'s amplitude starting at the black and ending at the red dotted lines in Fig.~\ref{Fig: DuffingInTime}. 

By fitting $f_i$ vs.~$A_i^2$ and $f_j$ vs.~$A_i^2$ during the ring ups to lines according to Eq.~\ref{Eq: Duffing Freq Shift}, we extract a slope corresponding to each ring up cycle. We average these approximately 20 slopes to obtain the values of $\gamma_{ii}^{SD}$ and $\gamma_{ji}^{XD}$. We then repeat the above process while toggling the drive of mode $j$ to obtain the two remaining coefficients (Table \ref{Table: Device Params}). We calibrate the Duffing coefficients in the same location on the trampoline as that used during the frequency tracking experiments.

\begin{figure}[t!]
    \centering
    \includegraphics{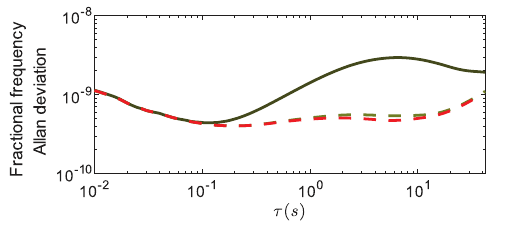}
    \caption{\small Testing the Duffing coefficient optimal for $\delta A$-$\delta f$ noise reduction. Common-mode subtracted Allan deviation (green solid line, identical to in Fig.~\ref{Fig: Subtracted ADevs with Theory}(b)) and corresponding Duffing correction utilizing (1) the measured coefficient as in Fig.~\ref{Fig: Duffing Correction} (green dashed line), and (2) values that empirically provide the best correction, ($\frac{\gamma^{SD'}_{aa}}{\gamma^{SD}_{aa}}, \frac{\gamma^{XD'}_{ab}}{\gamma^{XD}_{ab}}, \frac{\gamma^{XD'}_{ba}}{\gamma^{XD}_{ba}}, \frac{\gamma^{SD'}_{bb}}{\gamma^{SD}_{bb}})= (1.1, 0.6, 1.2, 0.7)$ (red dashed line).}
    \label{Fig: DuffingScan}
\end{figure}

In addition to measuring the Duffing coefficients, $\gamma$, experimentally, we also find that our measured Duffing coefficients are near optimal for the Duffing correction. We perform a sweep of each of the four Duffing coefficients, where the swept coefficients, $\gamma'$, are used to Duffing-correct the highest amplitude (darkest green) trace in Fig.~\ref{Fig: Subtracted ADevs with Theory}(b). Each Duffing coefficient is multiplied incrementally by a factor of 0.5 to 1.5 in steps of 0.1 such that all 11$^4$ possible combinations are tested. The efficacy of the resulting Duffing correction is evaluated by calculating the integral of the Duffing-corrected Allan deviation over $10^{-3}$ s $< \tau < 40$ s. Fig.~\ref{Fig: DuffingScan} shows the swept Duffing correction that yielded the lowest value of such integral (highest performance). Here, the factors used were $(\frac{\gamma^{SD'}_{aa}}{\gamma^{SD}_{aa}}, \frac{\gamma^{XD'}_{ab}}{\gamma^{XD}_{ab}}, \frac{\gamma^{XD'}_{ba}}{\gamma^{XD}_{ba}}, \frac{\gamma^{SD'}_{bb}}{\gamma^{SD}_{bb}})= (1.1, 0.6, 1.2, 0.7)$. Notably, this Duffing correction barely improves upon that obtained with the measured Duffing coefficients.

\section{Details of the Common-Mode Rejection}
\subsection{Correlated Frequency Drifts}
\label{Sec: Correlated Drifts}

\begin{figure}[t!]
    \centering
    \includegraphics{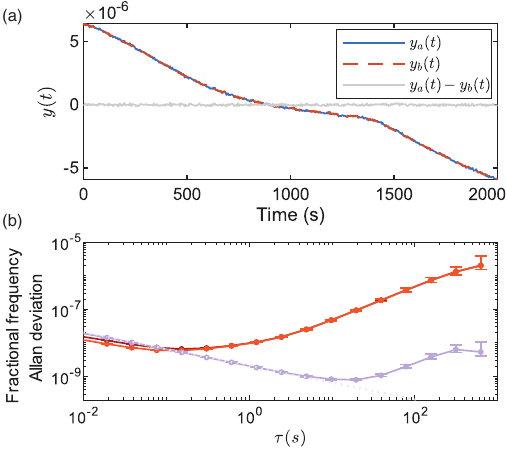}
    \caption{\small Long-term fractional frequency behavior. (a) Fractional frequency traces in time corresponding to the data shown in Fig.~\ref{Fig: Subtracted ADevs with Theory}(a) in main text ($y_a(t)$ in blue, $y_b(t)$ in orange, and $y_a(t)-y_b(t)$ in gray). (b) Allan deviations identical to those shown in Fig.~\ref{Fig: Subtracted ADevs with Theory}(a) but with averaging times past 50 s now shown. Error bars are 68\% confidence intervals, where a white frequency noise model was assumed given the white behavior of the common-mode subtraction and the presence of drifts at later times \cite{riley2008handbook}.}
    \label{Fig: DriftsInTime}
\end{figure}

Fig.~\ref{Fig: DriftsInTime} shows the long-term behavior of our fractional frequency data. In Fig.~\ref{Fig: DriftsInTime}(a), we see that drifts with a timescale around 1000 s exist, which matches closely with a $\sim$20-minute thermostat cycle observed in our laboratory. These drifts are correlated fractionally between the two modes (orange and blue traces) such that they are well subtracted out over long timescales (gray trace). Fig.~\ref{Fig: DriftsInTime}(b) shows the same data as in Fig.~\ref{Fig: Subtracted ADevs with Theory}(a) in the main text, but with $\tau > 50$ s shown. Not only does the common-mode subtraction technique reach the thermomechanical limit at $\tau < 10$ s, but it also suppresses single-mode noise by over two orders of magnitude at long times. Drift remaining in $y_a - y_b$ at long times are due to uncorrelated noise. This residual noise occurs beyond the timescale of Duffing noise\cite{manzaneque2023resolution}, and thus it does not hinder our Duffing correction method. That the common-mode subtracted Allan deviations reach thermomechanically-limited stability at times as large as 10 s is significant: recently such limit has been achieved but is restricted to $\tau < $ 0.1 s, or less \cite{piller2022thermal, martini2025uncooled, gavartin2013stabilization, sadeghi2020frequency}. 

We attribute the fact that we reach the thermomechanical limit at such long times to our system of a patterned, interferometrically-detected, low-optical absorption silicon nitride resonator, combined with the common-mode subtraction technique. We now enumerate these factors in detail. First, dissipation dilution \cite{engelsen2024ultrahigh} means that modes $a$ and $b$ of our trampoline~\cite{bereyhi2022perimeter, norte2016mechanical} have high quality factors exceeding 10$^6$. Thus, the modal linewidths are narrow, leading to high resolution of the thermomechanical noise above the detection noise floor. Second, the low optical scattering of Si$_3$N$_4$ allows for integration of our resonator with a low-noise optical detection system \cite{brubaker2022optomechanical, peterson2016laser, purdy2012cavity}. We employ balanced detection to ensure maximal cancellation of residual intensity noise of our YAG laser. At frequencies above 100 kHz, the imprecision noise floor of our detection laser system is a few fm$/\sqrt{\text{Hz}}$ for $\sim$3 mW of detection power incident on the resonator. We therefore achieve thermomechanical peaks that are resolved by a factor of seven and five orders of magnitude compared to the detection noise floor (Fig.~\ref{Fig: App Noise Injection}(b)), for modes $a$ and $b$, respectively. These conditions far exceed the criterion observed by Be\v{s}i\'c et al.~that a factor of 1000 is needed to observe the thermomechanical limit in an Allan deviation \cite{bevsic2023schemes}. Third, Si$_3$N$_4$ provides low optical absorption compared to low-stress SiN \cite{kanellopulos2024stress}, which helps evade photothermal backaction \cite{sadeghi2020frequency, kanellopulos2025comparative} originating from our detection and driving lasers.

\subsection{Choice of Modes Based on Other Trials of Common-Mode Rejection}
\label{Section: Other CMS}
Our choice of mechanical modes was dictated by which combination of modes yielded the lowest baseline instability via common-mode subtraction. We also chose two modes with comparatively high quality factors to achieve a low thermomechanical limit (Eq.~\ref{Eq: Thermomech limit}). This low instability would allow for the cleanest observation and thus correction of Duffing noise.

We tested different combinations of modes, e.g.~higher-order trampoline modes and torque modes, and found that the combination of the second- (mode $a$) and third- (mode $b$) higher-order symmetric trampoline modes yielded the best subtraction, with fractional frequency Allan deviations consistently reaching below $10^{-9}$ at high drive amplitude. Other mode combinations performed slightly worse, with the worst reaching just above $10^{-8.5}$ Allan deviation. 

We also tested common-mode subtraction on other devices with different geometries, from different Si$_3$N$_4$ deposition runs and using different relative drive parameters. We found that on other identical trampolines from the same deposition run as the device in the main text, we achieved the same low noise levels as reported there. We observed worse Allan deviations ($\sim 10^{-7}$) on an identical geometry trampoline but using Si$_3$N$_4$ that was observed under microscope to be less than pristine. Other geometries from that Si$_3$N$_4$ run also had higher measured Allan deviations. Currently, the sources of differences in performance among different mechanical modes and different fabrication iterations are unclear.

The empirical tests described in the previous paragraph support the intuition that noise affecting modes with similar symmetry (as in Fig.~\ref{Fig: Setup}(b)) would be the most correlated. Modes $a$ and $b$ also both have large mechanical motion at the central trampoline pad. This large motion lends itself to interferometric detection at this location since here our system is least affected by drifts of the probe beam relative to device features with sizes on the order of or smaller than the beam diameter.

\section{Radiation Pressure Drive}
\label{Section: Radiation Pressure Driving}
We drive our mechanics optically through radiation pressure achieved by amplitude modulating our 980 nm drive laser. This modulation is done by passing the light through an acousto-optic modulator (AOM) that is driven by 80 MHz RF, modulated with sinusoidal driving tones at the mode $a$ and $b$ frequencies output by the lock-in amplifier.

\subsection{Driving with White Noise}
\label{Section: white noise temperature}
In Fig.~\ref{Fig: Duffing Correction}(b), we add 2 MHz bandwidth white force noise to the coherent radiation pressure drive described above. This chosen bandwidth is much higher than the frequencies of modes $a$ and $b$ such that the noise is white in the vicinities of the modes. This experimentally applied noise is effectively equivalent to the scenario of a global thermal force due to the resonator's contact with a thermal bath at $T >> 295$ K \cite{manzaneque2023resolution}. 

To determine the effective temperature in the presence of added white noise, we fit the thermomechanical limit (Eq.~\ref{Eq: Thermomech limit} for our two modes added in quadrature) to our noise-added, common-mode subtracted Allan deviation data, where temperature $T$ is a free parameter and resonator parameters are those in Table \ref{Table: Device Params}. For this fit, we assume a perfectly calibrated interferometer response given that the laser wavelength is a well known parameter.

\begin{figure}[t!]
    \centering
    \includegraphics{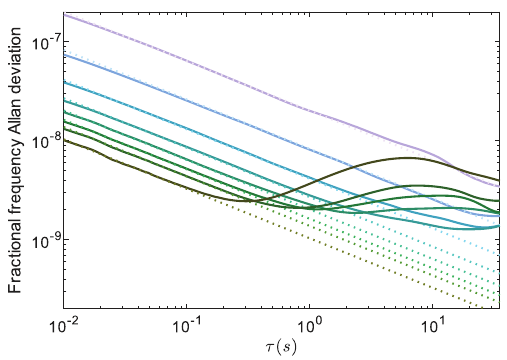}
    \caption{\small Common-mode subtracted Allan deviations (solid lines) with added white force noise for $\frac{A_a}{A_{\text{crit},a}^{\text{SD}}} = \frac{A_b}{A_{\text{crit},b}^{\text{SD}}} =  $ 0.14, 0.35, 0.71, 1.06, 1.41, 1.77, 2.12, 2.83, as in the main text. Dotted lines are the thermomechanical limit  for $T = 27,000$ K (Sec.~\ref{Section: white noise temperature}) at each drive.}
    \label{Fig: App Noise Injection}
\end{figure}

We find that the amount of white force we added corresponds to an effective temperature $T = 27,000$ K. Fig.~\ref{Fig: App Noise Injection} shows the thermomechanical limit for this temperature (dotted lines) as the drive amplitude is changed. We see good agreement with the data for all drive amplitudes. We thus use $T = 27,000$ K in all noise-added analysis in both the main text and in Sec.~\ref{Section: Amp Noise}.

\subsection{Radiation Pressure Estimate}
\label{Section: Radiation Pressure Estimate}
Here, we calculate the amount of radiation pressure used to drive our mechanics and compare the expected resulting displacement to the amplitude measured by the Michelson interferometer.

For the coherent drive, we use tens of $\mu$Ws of modulated power, which equates to forces on the order of $10^{-13}$ N, to drive our modes to tens of nanometers. We also add about $10^{-29}$ N$^2$/Hz of white force noise in the added noise case. Both the coherent and white measured radiation pressures yield coherent amplitudes $A_0$ and noise amplitudes $\sigma_{\text{th}}$, respectively, that match that expected from the Michelson interferometer measurement to within a factor of two or less.

\section{Amplitude Noise Contributing to $\delta A$-$\delta f$ Conversion}
\label{Section: Amp Noise}
Here, we report additional information on the nature of the amplitude noise, $\delta A$, that causes $\delta A$-$\delta f$ conversion in the nonlinear regime in our system. Given the nature of our Duffing correction technique, the main conclusions and capabilities of our method are independent of the character of this noise, but we provide this section for reference. 

\subsection{Characterizing the Amplitude Noise}
\label{Section: Temp Vs Drive}

For different drive amplitudes, we fit $A_{a}(t)$ and $A_{b}(t)$ to normal distributions and use the equipartition theorem to extract amplitude-noise-related temperatures for both the room temperature and white noise-added data. 

\begin{figure}[t!]
    \centering
    \includegraphics{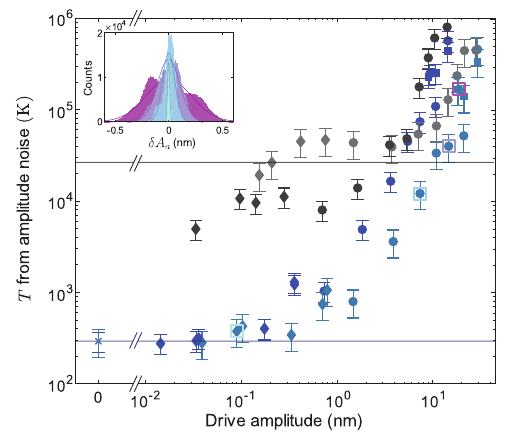}
    \caption{\small Examining the magnitude of the noise in the amplitude quadrature. Effective temperature versus drive amplitude of both modes assuming room temperature (mode $a$ light blue, mode $b$ dark blue) and added noise (mode $a$ light gray, mode $b$ dark gray). Data were taken on different days as indicated by differently shaped data points (x, diamond, circle, square). Blue horizontal line indicates 295 K and black horizontal line 27,000 K, as extracted from the free-parameter fit described in Sec.~\ref{Section: white noise temperature}. Inset: example histograms of raw mode $a$ amplitude noise data ($\delta A_a$) and normal distribution fits used to extract $T$. Different colors indicate different drive amplitudes corresponding to the points outlined in colored squares in the main plot.} 
    \label{Fig: TempVsDrive}
\end{figure}

For the room temperature case, Fig.~\ref{Fig: TempVsDrive} shows the extracted temperatures of the two modes in blue (light blue for mode $a$, and dark blue for mode $b$). The blue horizontal line indicates 295 K, the temperature of our laboratory. This data was taken on multiple different days, where each day is indicated by its own symbol (x, diamond, circle, or square).  Error bars represent the average percent difference between data points taken on different days at the same amplitude values. At zero drive, the magnitude of amplitude noise matches that expected from room temperature as shown by the points marked by x's in Fig.~\ref{Fig: TempVsDrive}. By contrast, we see that driving the modes harder adds a significant amount of noise such that at high drive, the extracted temperature is orders of magnitude greater than 295 K, and thus $\delta A > \Ath$. 

The gray points in Fig.~\ref{Fig: TempVsDrive} (light gray for mode $a$, dark gray for mode $b$) show equivalent data for when white force noise is added to our drive. The black line indicates the noise-added temperature, 27,000 K, described in Sec.~\ref{Section: white noise temperature}. We find that there is a wide overall spread in this data, but, on average, the temperature roughly matches expectation for a wider range of amplitudes than in the room temperature case, suggesting that in this regime $\delta A \approx \delta A_{\text{white}}$. However, at the highest amplitudes, the drive-dependent noise is still present such that $\delta A > \delta A_{\text{white}}$.

We also note that while we plot $T$ on the y-axis of Fig.~\ref{Fig: TempVsDrive} to show that $\delta A$ at high drive has a variance much larger than that expected from thermomechanical noise for both $T = 295$ K and 27,000 K, this extra noise does not have a Gaussian distribution, as demonstrated by the poor fits at higher drive in Fig.~\ref{Fig: TempVsDrive} inset. Indeed, deviations from simple expectations are likely due to added amplitude noise originating from a source without a white force noise character. Ultimately, we find that the amplitude noise present in our system is highly drive-dependent and less Gaussian the higher the drive. Nevertheless, this noise is successfully absent from our frequency data when our Duffing correction is applied.

\subsection{$\delta A$ contribution to Duffing $\delta A$-$\delta f$ conversion}
\label{Section: Duffing AM-FM}
We find that the $\delta A$-$\delta f$ conversion magnitude stemming from Duffing (seen at $\tau = \tau_{c,a}, \tau_{c,b} $ in the Allan deviations of Figs.~\ref{Fig: Subtracted ADevs with Theory}(b) and \ref{Fig: Duffing Correction}(a)) can be traced to the magnitude of the drive-dependent $\delta A$. To analyze this $\delta A$-$\delta f$ conversion, we model the Allan deviation at this timescale by simulating the Allan deviation of the following quantity: 

\begin{equation}
\frac{\langle f_a \rangle - f^{SD}_{aa}A_a^2 - f^{XD}_{ab}A_b^2}{\langle f_a - f^{SD}_{aa}
A_a^2 - f^{XD}_{ab}A_b^2\rangle} - \frac{\langle f_b\rangle - f^{SD}_{bb}A_b^2 - f^{XD}_{ba}A_a^2}{\langle f_a - f^{SD}_{bb}A_b^2 - f^{XD}_{ba}A_a^2\rangle}.
\label{Eq: Duffing expectation}
\end{equation}
We define this Allan deviation as $\sigyosubduff$. Note that the average of $f_a$ and $f_b$ in the numerators ensures that $\sigyosubduff$ represents only the frequency noise arising from $\delta f_{\beta}$, which dominates at this timescale. To simulate $\sigyosubduff$, we must simulate $A_a(t)$ and $A_b(t)$, with their respective noises, $\delta A$. Here, we assume that $\delta A = \delta A_{\text{th}}$ for the room temperature case and $\delta A = \delta A_{\text{white}}$ for the noise-added case.

We now compare the simulated $\sigyosubduff$ to the local maxima seen in the data in the regime of $\tau =\tau_{c,a}, \tau_{c,b}$. At room temperature and for high drive amplitudes, the measured Allan deviation due to Duffing noise is higher than our simulated expectation. For example, for the trace where $\frac{A_a}{A_{\text{crit},a}^{\text{SD}}} = \frac{A_b}{A_{\text{crit},b}^{\text{SD}}} =  $ 2.12 (second highest amplitude trace in Fig.~\ref{Fig: Duffing Correction}(a)), the measured Allan deviation is an order of magnitude higher than the simulated $\sigyosubduff$ at $\tau =\tau_{c,a}, \tau_{c,b}$. This discrepancy is consistent with the finding in Fig.~\ref{Fig: TempVsDrive}  that $\delta A > \Ath$. In the noise-added case, however, the measured data is only a factor of 2 higher than simulation for the same amplitude data trace (highest amplitude shown in Fig.~\ref{Fig: Duffing Correction}(b)), which is consistent with $\delta A \approx \delta A_{\text{white}}$.

\bibliography{References}

\end{document}